\documentclass[aps,prl,preprint,amsmath,amssymb,showpacs,preprintnumbers,superscriptaddress]{revtex4-2}
\usepackage{CJK}
\usepackage{graphicx}
\usepackage{dcolumn}
\usepackage{bm}
\usepackage{tabularx}
\usepackage{multirow}
\usepackage{lettrine}
\usepackage{color}

\begin{document}

\title{Relativistic configuration-interaction density functional theory: nonaxial effects on nuclear $\beta\beta$ decay}
\author{Yakun Wang}
\affiliation{State Key Laboratory of Nuclear Physics and Technology, School of Physics, Peking University, Beijing 100871, China}
\author{Pengwei Zhao}
\email{pwzhao@pku.edu.cn}
\affiliation{State Key Laboratory of Nuclear Physics and Technology, School of Physics, Peking University, Beijing 100871, China}
\author{Jie Meng}
\email{mengj@pku.edu.cn}
\affiliation{State Key Laboratory of Nuclear Physics and Technology, School of Physics, Peking University, Beijing 100871, China}
\date{\today}

\maketitle

Nuclear $\beta\beta$ decay is a second-order weak process in which an even-even nucleus decays to its neighboring even-even nucleus by turning two neutrons to two protons while emitting two electrons.
A nucleus undergoes such decay by two distinguishable modes, i.e., the two-neutrino $\beta\beta$ ($2\nu\beta\beta$) and the neutrinoless $\beta\beta$ ($0\nu\beta\beta$) decays.
The $2\nu\beta\beta$ decay is a lepton-number-conserving process and has been observed in a group of nuclei.
The yet unobserved $0\nu\beta\beta$ decay is a lepton-number-violating process and its observation would signal the violation of lepton number, establish the Majorana nature of neutrinos, and shed light on the matter dominance in the universe.
The $0\nu\beta\beta$ decay provides also an effective way to determine the neutrino mass scale and hierarchy.
Due to its great importance, the detection of the $0\nu\beta\beta$ decay has become the goal of several experimental projects worldwide~\cite{Agostini2023Rev.Mod.Phys.025002}.

One of the most stringent limits on the $0\nu\beta\beta$-decay half-life is given by the GERmanium Detector Array (GERDA) experiment with $T_{1/2} > 1.8 \times 10^{26}$ yr~\cite{Agostini2020Phys.Rev.Lett.252502}.
The next-generation Large Enriched Germanium Experiment for Neutrinoless $\beta\beta$ Decay (LEGEND)~\cite{Cattadori2021Universe314} and the China Dark matter EXperiment (CDEX) (see more details about CDEX at http://cdex.ep.tsinghua.edu.cn/info/PUBLIC\_DATA/1411) aim to boost the $0\nu\beta\beta$-decay half-life of $^{76}$Ge up to $10^{28}$ yr, which will probe the normal hierarchy region of the neutrino mass ordering and make $^{76}$Ge as one of the most promising candidates for the observation of the $0\nu\beta\beta$ decay.

Motivated by the ambitious experimental project, theoretical investigations for the $0\nu\beta\beta$ decay of $^{76}$Ge have been widely performed.
The key ingredient lies in calculating the nuclear matrix element (NME) which is crucial to extract the beyond standard model parameters from the experimental half-life.
Several models including the configuration-interaction shell model (SM), the quasiparticle random-phase approximation (QRPA), the triaxial projected shell model (TPSM), the interacting boson model (IBM), the generator coordinate method based on the SM interaction (GCM-SM), the GCM based on the relativistic and nonrelativistic density functional theories (GCM-CDFT and GCM-NRDFT), and the valence-space in-medium similarity renormalization group have been used to predict the $0\nu\beta\beta$-decay NME of $^{76}$Ge.
However, the predicted NMEs differ by a factor of three, causing an uncertainty of one order of magnitude in the half-life for a given value of the effective neutrino mass.

The $0\nu\beta\beta$-decay NME depends sensitively on the initial and final nuclear wavefunctions.
Therefore, testing the validity of nuclear wavefunctions from nuclear structure aspects and from other observed nuclear weak-decay modes is essential for a reliable prediction of NMEs.
Thanks to the progresses achieved experimentally, data are available in both aspects.
On the nuclear structure side, the low-lying spectra and the reduced $E2$ matrix elements of $^{76}$Ge and its $\beta\beta$-decay partner $^{76}$Se are recently measured~\cite{Henderson2019Phys.Rev.C054313,Ayangeakaa2019Phys.Rev.Lett.102501}.
In particular, the nonaxial nature, i.e., triaxiality of $^{76}$Ge and $^{76}$Se is revealed by applying the rotational invariant sum-rule method to the reduced $E2$ matrix elements.
On the nuclear weak-decay side, the $2\nu\beta\beta$-decay half-life of $^{76}$Ge has been measured by GERDA collaboration~\cite{Agostini2015EuropeanPhysicalJournalC}.
These new experimental results provide us opportunities to constrain the model calculations and benchmark the obtained wavefunctions.

The SM and the IBM have been employed to calculate the spectroscopic properties, $2\nu\beta\beta$ and $0\nu\beta\beta$ decays of $^{76}$Ge.
However, the SM calculations are performed within a limited model space, and some deformation driving orbits and spin-orbit partners relevant to nuclear $\beta\beta$ decays are not included.
The IBM calculations for the $2\nu\beta\beta$ decay are performed with closure approximation, which may not be suitable for the $2\nu\beta\beta$ decay.
In comparison, nuclear DFT describe nuclear properties within a full model space, and many-body correlations such as those induced by nuclear triaxiality are taken into account by symmetry breaking.
Although the nuclear DFT are designed for nuclear ground states, good description for nuclear spectroscopic properties and $\beta\beta$ decays can be achieved by proper extensions.

In this Short Communication, a novel method called Relativistic Configuration-interaction Density functional (ReCD) theory is established to predict the $0\nu\beta\beta$-decay NME of $^{76}$Ge.
The ReCD theory combines the advantages of the SM and nuclear DFT.
It starts from a well-defined relativistic density functional whose self-consistent solution is a state at the minimum of the potential energy surface and includes already important physics.
The additional correlations are considered with shell-model calculations in a configuration space built on top of the self-consistent solution.
The ReCD theory developed here allows a fully microscopic and self-consistent treatment of nuclear triaxiality and can calculate both even-even and odd-odd nuclei.
Therefore, triaxiality that is crucial for modeling the properties of $^{76}$Ge can now be considered within a full model space, and the spectroscopic properties, $2\nu\beta\beta$- and $0\nu\beta\beta$-decay NMEs of $^{76}$Ge are studied on the same footing.
The validity of the ReCD theory is examined by describing the energy spectra, $E2$ transition probabilities, and the $2\nu\beta\beta$-decay NME of $^{76}$Ge.
The importance of nuclear triaxiality for a high-precision prediction of the $0\nu\beta\beta$-decay NME and the corresponding implications to the next-generation $0\nu\beta\beta$-decay experiments will be highlighted.
The related theoretical framework and the numerical details can be found in the Supplementary materials.

\begin{figure*}[htbp]
  \centering
  \includegraphics[width=\textwidth]{Fig1.eps}
  \caption{(Color online) (a)--(f) The low-lying energy spectra and $E2$ transition probabilities (W.u.) for $^{76}$Ge and $^{76}$Se calculated by the ReCD theory with (Triaxial) and without (Axial) the inclusion of the triaxial deformation, in comparison with experimental data (Exp)~\cite{Ayangeakaa2019Phys.Rev.Lett.102501,Henderson2019Phys.Rev.C054313,Kavka1995NuclearPhysicsA177211}.
  (g)--(h) The potential energy surfaces of $0^+$ states in the $(\beta,\gamma)$ plane, with neighboring contour lines separated by 0.2 MeV.
  The stars denote positions of the minimum energy.}
  \label{Fig:Energy-spectra}
\end{figure*}
In the rightmost column of Fig.~\ref{Fig:Energy-spectra}, the potential energy surfaces of $0^+$ states for $^{76}$Ge and $^{76}$Se are shown.
Triaxial energy minima are clearly seen.
The calculated triaxial deformations $\gamma$ at the energy minima are respectively $22^\circ$ and $32^\circ$ for $^{76}$Ge and $^{76}$Se, which are consistent with the experimental $\gamma$ values, about $27^\circ$ for $^{76}$Ge and $24^\circ$ for $^{76}$Se.
The triaxiality plays important roles for reproducing the spectroscopic properties.
Without the consideration of triaxiality (Fig.~\ref{Fig:Energy-spectra}a and d), the observed $\gamma$ bands, i.e., those built on top of the $2_2^+$ states, cannot be reproduced, and the excitation energy of the $2_1^+$ state for $^{76}$Ge is overestimated while the one for $^{76}$Se is underestimated.
After including the triaxiality (Fig.~\ref{Fig:Energy-spectra}b and e), the predicted yrast bands and $\gamma$ bands are in good agreements with data, indicating that the description of underlying wavefunctions are significantly improved by the triaxial deformation.

\begin{figure}[htbp]
  \centering
  \includegraphics[width=\textwidth]{Fig2.eps}
  \caption{(Color online) (a) The $2\nu\beta\beta$-decay NMEs of $^{76}$Ge calculated by the ReCD theory with (red shaded area) and without (blue shaded area) the triaxial deformation as functions of excitation energies of intermediate $1^+$ states, in comparison with the data~\cite{Agostini2015EuropeanPhysicalJournalC}.
  The results calculated by the IBM~\cite{Barea2013Phys.Rev.C014315} and the SM~\cite{Senkov2016Phys.Rev.C044334,Kostensalo2022PhysicsLettersB137170,Caurier2012PhysicsLettersB6264} are also listed for comparison.
  (b) The $0\nu\beta\beta$-decay NMEs and contributions from each coupling channel calculated by the ReCD theory for $^{76}$Ge with and without the triaxial deformation.
  The results taking no account of the effect induced by the quasiparticle configuration mixing (marked as ``PCDFT", see text for more details) are also listed for comparison.
  (c) Comparison of the $0\nu\beta\beta$-decay NMEs from different model calculations, which include the ReCD theory and PCDFT in the present calculations, as well as SM~\cite{Senkov2016Phys.Rev.C044334}, GCM-CDFT~\cite{Yao2015Phys.Rev.C024316}, GCM-NRDFT~\cite{Rodriguez2010Phys.Rev.Lett.252503}, GCM-SM~\cite{Jiao2017Phys.Rev.C054310}, TPSM~\cite{Wang2021Phys.Rev.C014320}, QPRA~\cite{Fang2011Phys.Rev.C034320,Hyvaerinen2015Phys.Rev.C024613,Mustonen2013Phys.Rev.C064302}, and IBM~\cite{Barea2015Phys.Rev.C034304}.}
  \label{Fig:NME}
\end{figure}

In Fig.~\ref{Fig:NME}a, the $2\nu\beta\beta$-decay NME $M^{2\nu}$ as functions of excitation energies of intermediate $1^+$ states are depicted, in comparison with the data~\cite{Agostini2015EuropeanPhysicalJournalC} and those from the IBM~\cite{Barea2013Phys.Rev.C014315} and SM~\cite{Kostensalo2022PhysicsLettersB137170,Senkov2016Phys.Rev.C044334,Caurier2012PhysicsLettersB6264} calculations.
Due to the fact that the two-body currents are not included in the present calculations, quenching factors are introduced to the $M^{2\nu}$.
Without quenching, the $M^{2\nu}$ given by the SM calculations in Refs.~\cite{Senkov2016Phys.Rev.C044334}, \cite{Kostensalo2022PhysicsLettersB137170}, and \cite{Caurier2012PhysicsLettersB6264} are respectively 0.329, 0.333, and 0.322 MeV$^{-1}$, being slightly larger than the present results.
The quenching factors used in the literature range from $0.68$ to $0.77$ for nuclei in the mass region $A\sim70$.
These two values set the lower and upper limits of the present $M^{2\nu}$ and those from the SM and IBM calculations.

As shown in Fig.~\ref{Fig:NME}a, the $M^{2\nu}$ are overestimated by the SM and IBM calculations.
This might be partially attributed to their limited model space, which is so-called ``$jj44$" consisting of $0f_{5/2}$, $1p_{1/2}$, $1p_{3/2}$, and $0g_{9/2}$ valence orbits.
By analyzing the single particle levels near the Fermi surface in our calculations, it is found that $0f_{7/2}$, $1d_{5/2}$, and $0g_{7/2}$ orbits are crucial for driving the triaxiality, but they are out of the model space of the SM and IBM.
Enlargement of the model space by including these orbits may lower the $M^{2\nu}$.
Moreover, in the IBM calculation, going beyond the closure approximation may further reduce the $M^{2\nu}$.

The $M^{2\nu}$ given by the ReCD theory agree satisfactorily with the data.
The running sums show that main contributions of $M^{2\nu}$ come from the intermediate $1^+$ states with excitation energies lower than 7 MeV.
Moreover, the $M^{2\nu}$ is moderately influenced by the triaxiality.
The less pronounced triaxial effects can be understood from the $2\nu\beta\beta$-decay operator (see the Supplementary materials).
The $2\nu\beta\beta$-decay operator is independent on momentum $|\bm{q}|$ of virtual neutrino.
Therefore, the change of $2\nu\beta\beta$-decay NME by triaxiality depends not only on the overlap between wavefunctions but also the energy difference $E_n- (E_{\mathrm{i}}+E_{\mathrm{f}})/2$.
The inclusion of triaxiality enhances both the overlap between wavefunctions and the energy difference $E_n-(E_{\mathrm{i}}+E_{\mathrm{f}})/2$, which cancels the triaxial effects and leads to a moderate increase of the $2\nu\beta\beta$-decay NME by around 11\%.
Besides the $2\nu\beta\beta$-decay NME, the Gamow-Teller strength distribution for the decay process from $^{76}$Ge to $^{76}$As (GT$^-$) and that from $^{76}$Se to $^{76}$As (GT$^+$) have been also examined.
The predicted position of the GT$^-$ resonance locates at 13 MeV, and most of the GT$^+$ strengths come from the final states with excitation energy lower than 7.5 MeV.

The effect of triaxiality on the $0\nu\beta\beta$ decay of $^{76}$Ge can be seen in Fig.~\ref{Fig:NME}b, in which the total $0\nu\beta\beta$-decay NME and contributions from the vector (VV), axial-vector (AA), axial-vector and pseudoscalar (AP), pseudoscalar (PP), and weak-magnetism (MM) coupling channels are depicted.
In the calculation of the $0\nu\beta\beta$ decay, bare axial-vector coupling constant, i.e., $g_{\mathrm{A}} = 1.27$, is adopted.
In contrast to the $2\nu\beta\beta$ decay, the triaxiality is significant to the $0\nu\beta\beta$ decay of $^{76}$Ge.
The $M^{0\nu}$ is enhanced by a factor of two after considering the triaxial deformation.
Such enhancement can be clarified in terms of the difference of nuclear shapes between $^{76}$Ge and $^{76}$Se.
By assuming the axial symmetry, the initial and final nuclei are predicted to have distinct shapes, i.e., a prolate shape ($\gamma = 0^\circ$) for $^{76}$Ge while an oblate shape ($\gamma = 60^\circ$) for $^{76}$Se.
Such remarkable shape differences reduce the similarity of the wavefunctions and lead to a small $M^{0\nu}$.
After the inclusion of triaxiality, a new dimension associated with the triaxial degree of freedom is opened up in the nuclear shape space.
The shapes minimizing the energies of $^{76}$Ge and $^{76}$Se are moved to those with significant triaxial deformations (see Fig.~\ref{Fig:Energy-spectra}g and h).
Thus, the overlap between the wavefunctions of $^{76}$Ge and $^{76}$Se are notably increased, which results in the significant enhancement of the $M^{0\nu}$.

The significant enhancement of the $0\nu\beta\beta$-decay NME by the triaxiality has positive implications to the next-generation experiments.
According to the $0\nu\beta\beta$-decay half-life $[T^{0\nu}_{1/2}]^{-1} = G_{0\nu}|M^{0\nu}|^2\langle m_{\beta\beta}\rangle^2$, it is clear that a larger NME corresponds to a shorter decay half-life for a given effective neutrino mass $\langle m_{\beta\beta}\rangle$.
The goal of the next-generation LEGEND aims to probe the normal hierarchy region with $\langle m_{\beta\beta}\rangle$ being around 10 meV.
Therefore, the decay half-life should be boosted respectively to $1.27\times10^{28}$ yr and $4.47\times10^{28}$ yr based on the predicted NMEs with and without triaxiality.
Since the half-life sensitivity of the LEGEND scales linearly with the detector mass, the reduction of the predicted half-life by a factor around four suggests that less amount of the germanium is needed to achieve the goal of the next-generation experiments.

Fig.~\ref{Fig:NME}b shows also the $M^{0\nu}$ from calculations without the configuration mixing, which are termed as projected CDFT (PCDFT) calculations in the following.
The configuration space of the PCDFT consists of only the intrinsic ground state.
Differences of the $M^{0\nu}$ between the PCDFT and ReCD theory thus reveal the effect of the quasiparticle configuration mixing, which is not considered in the previous DFT-based GCM calculations.
By including the configuration mixing, the $M^{0\nu}$ with triaxiality is reduced from 6.36 to 5.92, while the one without triaxiality is enhanced from 2.42 to 3.16.

In Fig.~\ref{Fig:NME}c, the $M^{0\nu}$ predicted by the ReCD and PCDFT are compared with those from other model calculations.
The present $M^{0\nu}$ with triaxiality are much larger than those given by the SM, GCM-SM, and TPSM calculations performed in limited model spaces.
It is expected that enlarging their model spaces would enhance the corresponding $M^{0\nu}$.
The $M^{0\nu}$ in the IBM calculation is slightly smaller than the present results with triaxiality.
The IBM-based $M^{0\nu}$ may also increase after enlarging its model space.
Some differences are also seen between the present $M^{0\nu}$ with triaxiality and those in the QRPA, which may partially originate from the assumed spherical or axial symmetry in the QRPA calculations.
The $M^{0\nu}$ given by the ReCD theory with triaxiality is nearly the same as the result in the GCM-CDFT and slightly larger than the one in the GCM-NRDFT.
It is emphasised that the DFT-based GCM limited to the axial deformation cannot reproduce the spectroscopic properties of $^{76}$Ge and $^{76}$Se.
Recently, the importance of isoscalar pairing to the $M^{0\nu}$ has been suggested.
Including the isoscalar pairing in the ReCD theory may slightly reduce the $M^{0\nu}$.
Investigating the isoscalar pairing effects simultaneously on the spectroscopic properties, $2\nu\beta\beta$ and $0\nu\beta\beta$ decays is an interesting topic in the future.
Moreover, the predicted NMEs of double Fermi transition with and without triaxiality are respectively 0.837 and 0.170, and those of double Gamow-Teller transition with and without triaxiality are respectively 1.194 and 0.264.
The results indicate that isospin and spin-isospin symmetries are violated in the ReCD calculations.
These symmetries might be broken to different extents in different models, which may also influence the predicted $0\nu\beta\beta$-decay NMEs.

In summary, the ReCD theory, which combines the advantages of the SM and the CDFT, is established.
It allows a fully microscopic and self-consistent treatment of nuclear triaxiality and describes nuclear spectroscopic properties, $2\nu\beta\beta$ and $0\nu\beta\beta$ decays on the same footing.
The spectroscopic properties and the $2\nu\beta\beta$-decay NME of $^{76}$Ge are well reproduced, which provide confidence for the prediction of the $0\nu\beta\beta$-decay NME.
The effects of triaxiality on the NMEs are explored in a full model space for the first time.
It is found that the inclusion of the triaxiality of $^{76}$Ge and $^{76}$Se enhances the NME of the $0\nu\beta\beta$ decay significantly by a factor around two and, thus, the predicted half-life would be reduced by a factor around four.
The present results indicate that the goals of next-generation experiments searching for the $0\nu\beta\beta$ decay in $^{76}$Ge, namely boosting the decay half-life up to $10^{28}$ yr and reaching the normal hierarchy region of neutrino mass, can be achieved using less amount of the germanium.

{\bf Conflict of interest}

The authors declare that they have no conflict of interest.

{\bf  Acknowledgments}

This work was partly supported by the National Natural Science Foundation of China (12141501, 12105004, 12070131001, 11875075,  11935003, and 11975031), the National Key R\&D Program of China (2017YFE0116700 and 2018YFA0404400), the China Postdoctoral Science Foundation under Grant No. 2020M680183, the High-performance Computing Platform of Peking University, and State Key Laboratory of Nuclear Physics and Technology, Peking University.

{\bf Author contributions}

Yakun Wang, Pengwei Zhao, and Jie Meng conceived and designed the project.
Yakun Wang developed the theoretical framework and performed the calculations.
All authors contributed to the analysis of the results and to the writing of the manuscript.

\end{document}